# THE SHAPES OF DARK HALOS

HANS-WALTER RIX
*Institute for Advanced Study, Princeton, NJ 08540, USA*
*and*
*Max-Planck-Institut für Astrophysik, 85740 Garching, Germany*

## 1. Introduction

Galactic and extragalactic astronomers often mean different things by "halo"; the former mean the spheroidal distribution of metal poor stars around the Milky Way (MW), the latter mean the massive, dark component providing the large dynamically inferred mass in and around galaxies. This review will focus on massive, dark halos. Despite the title of this conference, we cannot restrict our attention to the shape of the MW halo: very little would be to say. We must turn to cosmological simulations and to observations of external disk galaxies in the hope of deriving a representative picture.

Extensive effort has gone into studying the radial mass profile of dark halos, $\rho = \rho(r)$. In comparison, the question of the shapes of the halo isodensity surfaces ("the shapes of halos") has received little attention. Even if the shapes of these surfaces were restricted to concentric ellipsoids, a full determination of the axis ratios and orientation angles at each radius would be beyond reach of observational tests for the foreseeable future. Therefore we must turn to simpler questions:
• Are halos approximately spherical ($a \sim b \sim c$), oblate ($a \sim b > c$) or triaxial ($a > b > c$)?
• Are the axis ratios, $a(r) : b(r) : c(r)$, a strong function of radius?
• Are the axes of the isodensity ellipsoids aligned as a function of radius? Is the symmetry plane of the halo aligned with the galaxy's disk?
• How can observations determine "characteristic" axis ratios?

Halo shapes are of interest chiefly for two reasons. First, in order to relate local properties in the MW [*e.g.* the halo mass density at the position of the Sun, $\rho(R_\odot, z = 0)$ ] to global properties [*e.g.* the rotation curve, $v_c(R)$ ], the halo flattening must be known. For a given value of $v_c(R_\odot)$ the value of $\rho(R_\odot, z=0)$ will be much higher, if the total distribution is highly flattened rather than spherical. Similarly, the fraction of the circular speed attributed to different components in "disk–halo" fits of other galaxies, can depend sensitively on the shape of the halo. Second, the flattening of dark halos may provide a crucial clue to the nature of the dark matter (DM). If the DM were dissipational (e.g. baryonic MACHOs, cold molecular gas), then it should be distributed in disk-like structures. This is not the case for models with dissipationless DM.

## 2. Theoretical Expectations

What shape is predicted for the halos arising in hierarchical structure formation scenarios? For dissipationless DM this question can be addressed by cosmological N-body simulations (Katz, 1991; Dubinski and Carlberg, 1991; Warren *et al.*, 1992; Katz and Gunn, 1992; Summers, 1993; Dubinski, 1994). Competing requirements make even collisionless calculations difficult: on the one hand, they should be "cosmologically correct", i.e. should properly simulate the generation of angular momentum and the merging history for each halo. This requires large simulation volumes. On the other hand, the simulations must resolve galactic scales, $\lesssim 5$ kpc, in order to make predictions in a radial range where observational tests are feasible.

Despite an ongoing discussion about numerical techniques and proper initial conditions, all *dissipationless* simulations agree on a number of important results: (a) The resulting halos are strongly triaxial, closer to prolate than oblate. (b) The net angular momentum is aligned with the shortest axis. (c) In the virialized parts ($\lesssim 50$ kpc) the isodensity ellipsoids are aligned as a function of radius. (d) The axis ratios of the isodensity ellipsoids vary only weakly with radius, and the distribution of halo flattenings, $(c/a)_{halo}$, peaks near 0.7.

Dissipational material will affect the halo shape in several ways. A mass concentration at small radii will round the equipotentials by increasing the monopole. The concentration of mass towards a (disk) plane, will adiabatically compress the halo in the disk normal direction. Further, dissipative matter will settle onto loop orbits in a disk. These loop orbits will be elongated *perpendicularly* to the imposed distortion of the halo, and make the total (disk + halo) potential closer to axisymmetric than it would be due to either component (see Section 3).

Simulation including gas dynamics (e.g. Katz and Gunn 1992; Summers, 1993) or the slow growth of a disk component (Dubinski, 1994), bear out these expectation: the halos become slightly flatter and much closer to oblate ($b/a \sim 1$). The surprising result of the simulations is that these changes in halo shape extend to large radii, where $M_{dissipative}/M_{total} \ll 1$.

Hence the reshaping cannot be solely attributed to adiabatic orbit changes, but must be due to a decrease in the abundance of the box orbits, which originally sustained the triaxial shape.

## 3. Observational Signatures of Non-Spherical Halos

Rotation curves measure the monopole structure of the mass distribution, while the halo shapes are described by the higher mass multi-poles. Observational tests can only measure the shape of the *potential*, which is much rounder than the generating *mass distribution* and which has correspondingly smaller multipole terms. Therefore it should come as no surprise if shape measurements proved much more difficult than rotation curve measurements.

Most observational tests designed to measure the halo shapes are based on the following ideas:
(1) estimating the MW halo flattening through dynamical modelling of collisionless tracers. As a first step, one identifies a population of kinematic tracers that probe the potential at large distances and high above the disk, such as "extreme Population II" stars. Then one obtains the shape of the *local* velocity dispersion ellipsoid, the mean radial density profile and the flattening for the tracer population (not the total mass!). With these data and the run of the (equatorial) rotation curve, one can ask which flattening of the mass distribution, $(c/a)_{mass}$, is consistent with all these constraints.
(2) determining the shape of closed loop orbits. In any given potential there is a unique set of *closed loop orbits*. For a specific example, consider a non-rotating, logarithmic potential of constant distortion, $\epsilon_{pot}$,

$$\Phi(R,\varphi) \equiv \frac{v_c^2}{2}\left(\ln R^2 - \frac{\epsilon_{pot}}{2}\cos 2\varphi\right), \quad (1)$$

where $\epsilon_{pot} \sim \frac{1}{3}(1-(b/a)_{halo})$. In this potential the closed loops are simple ovals of the same ellipticity, $\epsilon_{pot}$, which are elongated *perpendicularly* to the equipotentials:

$$R(\varphi) = R_0(1 - \frac{\epsilon_{pot}}{2}\cos 2\varphi). \quad (2)$$

The velocities, $(v_\varphi, v_R)$, of any particle moving along these orbits differ from the circular motion $(v_c, 0)$ by a fractional amplitude of $2\epsilon_{pot}$. This very restricted set of orbits acquires its importance, because it constitutes the only orbits on which gas can move without getting shocked by other gas elements in the galaxy. Hence cold $(T \ll T_{virial})$ gas will populate these orbits after a few dynamical periods (Rix and Katz, 1991). Note that material on such loop orbits (Eq. 2) will produce a potential whose quadrupole distortion has the opposite sign of the input distortion (Eq. 1). In this sense the response of the embedded disk "counter-acts" the imposed halo distortion. Whenever disk and halo are of comparable dynamical importance, their combined potential will be much closer to axisymmetric than it would have been due to either component.

Other observational methods to measure the shape of halos include:
(3) studying the shape of the X-ray gas in halos. This technique has been applied to measuring the shape of several galaxy clusters (Buote and Canizares, 1992) and of one individual elliptical galaxy, NGC 720 (Buote and Canizares, 1994). If the X-ray gas is in quasi-hydrostatic equilibrium and if it is nearly isothermal, then the map of the projected gas emissivity provides a good map of the projected potential. However, it is unlikely that this method can be extended to halos around disk galaxies, since those do not seem to have smooth X-ray coronae.
(4) studying the flaring of neutral hydrogen disks (e.g. Maloney, 1993; Olling and van Gorkom, *in prep.*). The neutral hydrogen disks around spiral galaxies are observed to have a universal velocity dispersion of $\sigma \sim 8$km/s. In edge-on disks a comparison of the rotation curve (radial centrifugal equilibrium) to the vertical gas scale height (vertical hydrostatic equilibrium) can constrain the shape of the halo. In such an analysis only the potential in the immediate disk vicinity is probed and warps may be difficult to untangle from flaring. Nonetheless, this method may eventually test whether the DM at large radii is in a disk-like configuration.

## 4. Are Halos Flattened?

### 4.1. MILKY WAY

All existing estimates for the flattening of the dark MW halo arise from the dynamical modelling of extreme Population II stars (see Section 3). The observational constraints are: $(\sigma_R : \sigma_\phi : \sigma_z)_{trc} \sim 1.5 : 1 : 1$ at the position of the sun (Hartwick 1983, Sommer-Larsen and Zhen 1990, Morrison *et al.* 1990), $(c/a)_{trc} = 0.6 - 0.85$ (Wyse and Gilmore, 1986, Sommer-Larsen and Zhen 1990), $\rho_{trc} \propto R^{-3.5}$ and $v_c(R) = const$.

This observational input, however, is far from determining a unique model and further "plausible" assumptions need to be made. Binney, May and Ostriker (1987) constructed approximately self-consistent disk halo models with Stäckel potentials and find that $(c/a)_{halo} \sim 0.5$ fits the observations best. For a sample of local halo stars with known space motions, Sommer-Larsen and Zhen (1990) retraced the orbits in a Stäckel potential, favoring a nearly spherical halo, $(c/a)_{halo} \sim 0.9$. Van der Marel (1991) used the Jeans equations to show that halo flattenings of $0.4 \leq (c/a)_{halo} \leq 1$ were consistent with the data, depending on the tilt of the velocity el-lipsoid away from the disk plane. Amendt and Cuddeford (1994) derived higher order closure relations for the Jeans equations and applied those to Hawkin's (1984) sample of RR Lyrae stars, finding a preferred value of $(c/a)_{halo} \sim 0.7$.

In summary, these tests show that the data permit a wide range of halo flattenings, with only very flat halos disfavored by the data. It is unlikely that these tests can be much improved in the foreseeable future, because the dominant source of uncertainties is our ignorance about the shapes of the stellar tracer orbits.

### 4.2. POLAR RING GALAXIES

Polar ring galaxies are a relatively rare species of disk galaxies in which a ring of stars and gas orbits over the galaxy's pole. In such polar rings of varying radial extent, $0.1 \lesssim \Delta R/\langle R \rangle \lesssim 2$, the neutral and ionized gas provide an excellent probe of the halo flattening: (a) the gas moves in a plane including the $z$-axis, and (b) the gas often extends to radii where the dark halo is expected to dominate. The halo flattening can be estimated either by comparing the kinematics in the polar and equatorial (disk) plane or by studying the shape of the (closed loop) orbits in the polar plane. Sackett and Sparke (1990) and Sackett *et al.* (1994) built global dynamical models for NGC4650A to match simultaneously the kinematics of the stellar disk and of the polar ring and found $(c/a)_{halo} \sim 0.4 \pm 0.1$. For A0136-0801, where H$\alpha$ observations provide a high quality, two dimensional velocity field, Sackett and Pogge (*in prep.*) show that the halo must also be oblate. In UGC 7576 (Sackett, Rix and Peletier, *in prep.*), the ionized gas appears to form a thin ($\Delta R/\langle R \rangle \ll 1$), edge-on annulus. If gas is on circular orbits (in a spherical potential) then the line-of-sight velocities and the projected positions are related by $v_{l.o.s.}(R_{proj}) \propto R_{proj}$. If the potential is oblate, the closed gas orbits are oval and this relation becomes non-linear in a unique way. The gas velocities along the polar ring in UGC 7576 do indeed show such non-linearities and a very preliminary analysis points to flattening of $(c/a)_{mass} \sim 1/2$ at the radius of the ring.

In summary, the three disk galaxies with polar rings studied to date all show that their halos are flattened perpendicular to the stellar disk, with characteristic flattenings of $(c/a)_{halo} \sim 0.5$.

## 5. Are Halos Axisymmetric?

There are no direct constraints on deviations of the MW *halo* from axisymmetry in the disk plane. In part, this is due to the partial compensation of disk and halo distortions, discussed in Section 3. Even for significantly triaxial halos this will make the combined disk-halo potential much rounder than it would be due to either of its components, as long as the disk and the halo contribute comparably to the rotational support at the solar radius.

To address the axisymmetry of halos we must turn to external galaxies and rely on the Copernican principle for any inference about the MW. All existing observational tests for external disk galaxies are based on the shape of closed loop orbits. Franx and de Zeeuw (1993, FZ93) proposed a very powerful statistical way to limit $(b/a)_{pot}$, using the small scatter in the linewidth-luminosity relation ("Tully-Fisher" relation) for spiral galaxies. If the HI disks in these galaxies are oval then the same (edge-on) galaxy seen at different azimuthal angles, $\varphi$, will have linewidths, $W_{20}(\varphi)$, varying by $(1 + \epsilon_{pot} \cos 2\varphi)$. Averaged over all viewing angles the line width scatter, $\Delta W_{20}$ will be at least $\Delta W_{20}/W_{20} = \epsilon_{pot}/\sqrt{2}$. If the luminosity, $L$ varies as $L \propto (W_{20})^\gamma$, with $\gamma \sim 4$, this will result in a luminosity scatter of $\Delta L/L = \gamma \epsilon_{pot}/\sqrt{2}$. Using the observed scatter in published galaxy samples, Franx and de Zeeuw derive $\epsilon_{pot} < 0.1$. Unfortunately, it is unclear whether $W_{20}$ probes a radius where the halo dominates. This problem could be, and should be, remedied by considering the full two-dimensional velocity field in a small subsample of galaxies rather than only the HI linewidth.

Alternatively, one can use the shape of stellar disk isophotes to constrain the potential distortion in the disk plane. Even though stars to not move on *closed* loop orbits an extension of this analysis is possible as long as most stars move on loop orbits at all (Rix and Zaritsky, 1994). If an exponential disk with scale length $R_{exp}$ is embedded in a logarithmic potential, its isophote shapes, $\epsilon_{iso}$, are related to the potential shape by $\epsilon_{iso}(R) = \epsilon_{pot}(1 + R_{exp}/R)$ (FZ93). This relation can be applied statistically (Binney and de Vaucouleurs, 1981; Huizinga and van Albada, 1992; Fasano *et al.* 1993, FZ93), comparing the distribution of apparent isophotal axis ratios of disk galaxies to the expected uniform distribution expected for randomly oriented, circular disks. All these authors all find $\langle \epsilon_{pot} \rangle \lesssim 0.1$.

Rix and Zaritsky (1994) tried to improve on these analyses by using near-infrared K(2.2$\mu m$) images (reducing the distorting impact of dust and hot stars on the isophote shapes), by selecting the sample galaxies to be face-on on a purely kinematic basis and by fitting $\epsilon_{iso}(R)$ over the whole disk rather than measuring the disk ellipticity at one single radius. They found a tighter limit for the characteristic potential ellipticity, $\langle \epsilon_{pot} \rangle \lesssim 0.05$ for $R_{exp} \leq R \leq 4R_{exp}$. But even this measurement constrains $(b/a)_{halo}$ only weakly. If disk and halo contribute comparably to $v_c$, then their quadrupole distortions will cancel in part and $\langle \epsilon_{pot}^{halo} \rangle \sim 2\langle \epsilon_{pot} \rangle$ (Dubinski 1994). This makes $(b/a)_{halo} \sim 1 - 3\langle \epsilon_{pot}^{halo} \rangle \sim 0.7$ still consistent with the data.

The most compelling constraint on the shape of any halo exists for the elliptical galaxy IC 2006 (Franx, van Gorkom and de Zeeuw, 1994): $\epsilon_{pot} = 0.015 \pm 0.025$ at $R = 7R_e$. Since the halo shape seems to be affected by the orbital structure of the luminous mass (the disk-halo "counter-action"), it is not clear whether the halo shapes in elliptical and disk galaxies should be the same.

## 6. So, What is the Shape of Dark Halos?

Cosmological simulations of halo formation predict that they are flattened by $\langle c/a \rangle \sim 0.7$, and that their axis ratios $\langle b/a \rangle$ depend sensitively on the amount of dissipative matter in the halo.

Our position within the MW complicates measurements of the halo shape for the MW. Consequently, a wide range of shapes is consistent with the data. Some progress has been made on constraining the shapes of halos in external disk galaxies. In particular, flattening estimates are now becoming available for several polar ring galaxies. In these galaxies the halos are definitely flattened perpendicular to the stellar disk (with $(c/a)_{halo} \sim 0.5$), consistent with cosmological formation simulations of dissipationless DM halos. Some more work would be desirable to see whether very flat halos (e.g. from dissipational DM) can be ruled out observationally. Constraining deviations of the halo from axisymmetry in the disk plane, is complicated by the counter-action of halo and disk distortions. This problem is best remedied by going to large radii, where the halo dominates (see e.g. the tight constraint on the shape of the halo around the elliptical galaxy IC2006; Franx, van Gorkom and de Zeeuw, 1994).

If the MW halo is typical, then observations of the halos around other galaxies and the cosmological simulations lead us to adopt an axisymmetric halo with $(c/a) \sim 1/2$ as a better "default" shape than spherical when modelling the MW.

It is a pleasure to thank Penny Sackett and E. Q. Reade for extensive help in preparing this review.

## References

Amendt, P. and Cuddeford, P., 1994, *preprint*.
Binney, J., May, A. and Ostriker, J. P., 1987, MNRAS, **226**, 149.
Binney, J. and de Vaucouleurs, G., 1981, MNRAS, **194**, 679.
Buote, D. and Canizares, C., 1992 ApJ, **400**, 385.
Buote, D. and Canizares, C., 1994 ApJ, **427**, 86.
Dubinski, J. and Carlberg, R., 1991, Ap.J., **369**, 13.
Dubinski, J., 1994, *preprint*.
Dubinski, J. and Kuijken, K., 1994, *preprint*.
Fasano, G. *et al.*, 1993, MNRAS, **262**, 109.
Franx, M. and de Zeeuw, T., 1992, ApJL, **392**, 47.
Franx, M., van Gorkom, J. and de Zeeuw, T., 1994, ApJ, *in press*.
Hartwick, F. D. A., 1983, MmSocAstIt, **54**,51.
Hawkins, M. R. S., 1984, MNRAS, **206**, 433.
Huizinga, J.E. and van Albada, T.S., 1992, MNRAS, **254**, 677.

Katz, N., 1991, Ap.J.,**368**, 325.
Katz, N. and Gunn, J. E., 1991, Ap.J.,**377**, 365.
Maloney, P., 1993, ApJ,**414**, 41.
Morrison, H., Flynn, C. and Freeman, K., 1990, A.J.,**100**, 1191.
Rix, H.-W. and Katz, N., 1991, in *Polar Rings and Inclined Disks in Galaxies*, Cambridge Univ. Press, Eds: S. Casertano, P. Sackett and F. Briggs.
Rix, H.-W. and Zaritsky, D. F., 1994, Ap.J. *submitted*.
Sackett, P. and Sparke, L., 1990, ApJ, **361** 408.
Sackett, P., Rix, H.-W., Jarvis, B. and Freeman, K., 1994, Ap.J., *in press*.
Sommer-Larsen, J. and Zhen, C., 1990, MNRAS, **242**, 10.
Summers, F., 1993, Ph.D. Thesis, Univ. of California, Berkeley.
van der Marel, R. P., 1991, MNRAS, **248**, 515.
Warren, M. S., Quinn, P. J., Salmon, J. K. and Zurek, W. H., 1992, Ap.J., **399**, 405.
Wyse, R. F. G, and Gilmore, G., 1986, A.J., **91**, 855.